# AN EMBEDDING FOR GENERAL RELATIVITY AND ITS IMPLICATIONS FOR NEW PHYSICS

Bahram Mashhoon [1,3] and Paul Wesson [2,3]

1. Dept. Physics and Astronomy, University of Missouri, Columbia, Missouri 65211, U.S.A.

2. Dept. Physics and Astronomy, University of Waterloo, Waterloo, Ontario N2L 3G1, Canada

3. The S.T.M. Consortium, http://astro.uwaterloo.ca/~wesson

Correspondence: Mail to addresses 1,2 above; emails = mashhoonb@missouri.edu, wesson@astro.uwaterloo.ca (or pswastro@hotmail.com - not for publication)

Abstract

We show that any solution of the 4D Einstein equations of general relativity in vacuum with a cosmological constant may be embedded in a solution of the 5D Ricci-flat equations with an effective 4D cosmological "constant" Λ that is a specific function of the extra coordinate. For unified theories of the forces in higher dimensions, this has major physical implications.

1. Introduction

The embedding of Einstein's four-dimensional equations of general relativity in a higher-dimensional space provides a promising route to the unification of gravity with the interactions of particle physics. Five dimensions is of particular interest, since it represents the simplest extension of spacetime and is widely regarded as the low-energy limit of even higher-dimensional theories with relevance to particles, such as 10D supersymmetry, 11D supergravity and higher-D versions of string theory. In the present account, we present a 4D/5D embedding theorem with wide implications for the cosmological "constant" problem, the Weak Equivalence Principle and the evolution of the universe.

Embedding theorems are currently of interest because they constrain the ways in which classical general relativity can be welded to higher-D manifolds which may embody the internal symmetry groups characteristic of particles. For example, Campbell's theorem as recently resurrected by Tavakol and coworkers, provides a kind of ladder to go between manifolds whose dimensionality differs by one [1-5]. At the low end of this ladder, it implies that the 4D Einstein equations may be embedded in the 5D Ricci-flat equations (i.e., those where the Ricci tensor, rather than the Riemann-Christoffel curvature tensor, is set to zero). This provides an algebraic basis to physical accounts, such as membrane theory and induced-matter theory [6,7], which are now known to be mathematically equivalent [8,9]. However, the calculation of physical effects in such theories still requires the specification of a 5D line element, whose metric coefficients represent the potentials, as in the warp metric or the canonical metric [10]. Below, we will use the latter, since it simplifies the algebra. It should be noted, in this regard, that several studies have been made of the generality of this 5D metric [11-14]. It is analogous to the synchronous metric used in standard 4D cosmology [15-17]. Below we will study its most general form. We will thereby include results found recently on the embedding of 4D conformally-flat cosmologies of the de Sitter type which are relevant to the early inflationary period of the universe [14]. Our work will also include an exact 5D solution found recently which embeds the 4D Schwarzschild - de Sitter solution and is relevant to the solar system [5]. Further, we will obtain exact results on the value of the cosmological "constant" which is derived when a 5D metric is reduced to a 4D one [18-20]. This provides extra insight to the mismatch of this parameter as calculated in macroscopic and microscopic situations [21-28], which may be related to the different 4D physics that follows from different choices of 5D gauge [29]. Our results will also confirm the inference that the Weak Equivalence Principle, which ensures the equality of accelerations in a gravitational field for objects of all types, can be viewed as the 4D dynamical consequence of a 5D geometric symmetry [30; see also 11, 31, 32 for an extra force which can arise from the fifth dimension; and 12, 33 for the motion of massive particles along timelike paths in 4D which can be null paths in 5D]. In summary, we will recapture a large number of recent results from our theorem, while opening up new routes of investigation.

Our embedding theorem, which is in essence a special case of that of Campbell but focussed on physical effects, is proved algebraically in Section 2. Its consequences – such as exact 5D solutions – may be verified numerically using a fast computer package, for example GRTen-sor. Those readers who are more interested in physics than mathematics may like to proceed to Section 3, where we review our results and discuss their implications for new physics. These include changes in the cosmological "constant" and violations of the Weak Equivalence Principle, which may be tested using galaxies and artificial satellites [34, 35]. In these and other ways, we can test if the world has more than four dimensions.

2. The 4D Vacuum Einstein Equations as a Subset of the 5D Ricci-Flat Equations: An Embedding Theorem for Spacetime

In this section, we absorb the speed of light $c$ and the gravitational constant $G$ via a choice of units which renders them unity. We let lower-case Greek letters run 0,123 for time ($t$) and space ($xyz$ or $r\theta\phi$), defining a 4D proper time or interval in terms of a metric tensor via $ds^2 = g_{\alpha\beta}dx^\alpha dx^\beta$ (with summation over repeated indices as usual). This is part of a 5D interval defined analogously via $dS^2 = g_{AB}dx^A dx^B$, where upper-case Latin indices run 0,123,4 and the extra coordinate is $x^4 \equiv l$. Our goal is to isolate the most useful form for the 5D line element, given the restrictions implicit in the field equations. In this way, we will derive a special form of Campbell's theorem, which however is physically general and allows us to see how the fifth dimension constrains what happens in 4D spacetime.

The 4D Einstein field equations of general relativity are commonly written as $G_{\alpha\beta} + \Lambda g_{\alpha\beta} = 8\pi T_{\alpha\beta}$. Here, $G_{\alpha\beta} \equiv R_{\alpha\beta} - Rg_{\alpha\beta}/2$ is the Einstein tensor in terms of the Ricci tensor and its scalar, $\Lambda$ is the cosmological constant, and $T_{\alpha\beta}$ is the energy-momentum tensor which contains the material (non-vacuum) sources. However, the latter are often negligible in astrophysical situations. Then the field equations take the form

$$R_{\alpha\beta} = \Lambda g_{\alpha\beta} \quad (\alpha, \beta = 0,123) \quad , \tag{1}$$

defining an Einstein space. These equations, with appropriate sources and a small or zero value for $\Lambda$, are known to provide a good description of the dynamics of the solar system and other astrophysical systems such as binary pulsars. Their analog in five dimensions is

$$R_{AB} = 0 \quad (A, B = 0,123, 4), \tag{2}$$

defining a Ricci-flat space. These equations are often used to provide a unified description of the gravitational, electromagnetic and scalar interactions in a classical sense. Alternatively, they describe the interactions associated with a spin-2 graviton, a spin-1 photon and a spin-0 scaleron.

It is already known that equations (2) contain equations (1), by virtue of Campbell's theorem. However, to bring out the physics inherent in (2), we need to assume a form for the 5D metric. Following earlier work, we choose to use the five available degrees of coordinate freedom to suppress the potentials of electromagnetic type $(g_{4\alpha} = 0)$ and flatten the potential of scalar type $(g_{44} = -1)$. The latter condition means that our coordinate system is analogous to the synchronous one of general relativity (though with $l$ replacing $t$): the $l$-lines are the congruence of geodesics normal to 4D spacetime, and all observers agree on the value of $x^4 = l$. The 5D metric thus has the form $dS^2 = ds^2 - dl^2$, where the spacetime part depends in general on $x^\gamma$ and $x^4 = l$. (We will here and below make the usual assumption that the extra dimension is space-like.) However, again following earlier work, it proves useful to factorize the 4D part of the 5D metric by an $l^2$ term. There are three reasons for this: (1) we can make contact with the large literature which already exists on 5D metrics of so-called canonical type (see below); (2) we gain insight by comparison with the Milne model of general relativity (whose 3D part is modulated by an analogous factor of $t^2$); (3) we may consider the resulting metric as one involving a kind of moment in the 2-plane, since $dS^2 = ds^2 - dl^2$ under $s \to l \sin h(s/L)$ and $l \to l \cos h(s/L)$ becomes $dS^2 = (l/L)^2 ds^2 - dl^2$. Here $L$ is a constant length introduced for the consistency of physical dimensions and whose meaning will become clear below. Then we obtain

$$dS^2 = (l/L)^2 g_{\alpha\beta}(x^\gamma, l) dx^\alpha dx^\beta - dl^2 \quad . \tag{3}$$

This is the canonical form, which aids understanding in a physical sense but is still general in a mathematical sense.

The field equations (2) for metric (3) involve components of the 5D Ricci tensor which are relatively simple in form. They are:

$$^{(5)}R_{44} = -\frac{\partial A^\alpha{}_\alpha}{\partial l} - \frac{2}{l} A^\alpha{}_\alpha - A_{\alpha\beta} A^{\alpha\beta} \quad , \tag{4a}$$

$$^{(5)}R_{\mu 4} = A^\alpha{}_{\mu;\alpha} - \frac{\partial \Gamma^\alpha{}_{\mu\alpha}}{\partial l} \quad , \tag{4b}$$

$$^{(5)}R_{\mu\nu} = {}^{(4)}R_{\mu\nu} - S_{\mu\nu} \quad , \tag{4c}$$

where $S_{\mu\nu}$ is a symmetric tensor given by

$$S_{\mu\nu} \equiv \frac{l^2}{L^2}\left[ \frac{\partial A_{\mu\nu}}{\partial l} + \left(\frac{4}{l} + A^\alpha{}_\alpha\right) A_{\mu\nu} - 2 A_\mu{}^\alpha A_{\nu\alpha} \right] + \frac{1}{L^2}\left(3 + l A^\alpha{}_\alpha\right) g_{\mu\nu} \quad . \tag{5}$$

Here $^{(4)}R_{\mu\nu}$ and $\Gamma^{\mu}_{\nu\rho}$ are, respectively, the 4D Ricci tensor and the connection coefficients constructed from $g_{\alpha\beta}$. Moreover

$$A_{\alpha\beta} \equiv \frac{1}{2}\frac{\partial g_{\alpha\beta}}{\partial l} \quad , \tag{6}$$

where $A_{\alpha}^{\beta} = g^{\beta\delta}A_{\alpha\delta}$, and the semicolon in equation (4b) represents the usual 4D covariant derivative. We need to solve (4) in the form $R_{AB} = 0$.

This problem has been considered by various workers in the past under different motivations. Our motive is the embedding of 4D in 5D, so we now split the general metric tensor in (3) into two parts via $g_{\alpha\beta}(x^{\gamma},l) \equiv \chi(x^{\gamma},l)g^{*}_{\alpha\beta}(x^{\gamma})$. In solving $R_{AB} = 0$ subject to this split, it is convenient to tackle equations (4) by turn. The working is tedious, so we only note the main points in the following three paragraphs.

In (4a), we have $A_{\alpha\beta} = \chi' g^{*}_{\alpha\beta}/2$, where $\chi' \equiv \partial\chi/\partial l$, so $A^{\alpha\beta} = \chi' g^{*\alpha\beta}/(2\chi)^2$, $A^{\alpha}_{\beta} = \chi'\delta^{\alpha}_{\beta}/(2\chi)$, $A^{\alpha}_{\alpha} = 2\chi'/\chi$ and $A_{\alpha\beta}A^{\alpha\beta} = (\chi'/\chi)^2$. The noted equation set to zero then gives $\partial A^{\alpha}_{\alpha}/\partial l + 2A^{\alpha}_{\alpha}/l + A_{\alpha\beta}A^{\alpha\beta} = 0$ or $2\partial(\chi'/\chi)\partial l + 4\chi'/(l\chi) + (\chi'/\chi)^2 = 0$. The general solution of this is

$$\chi(x^{\gamma},l) = \left[1 - \frac{l_0(x^{\gamma})}{l}\right]^2 k(x^{\gamma}) \quad , \tag{7}$$

where $l_0(x^\gamma)$ is an arbitrary length which we will find in the next paragraph is constant, while $k(x^\gamma)$ is another arbitrary function of integration.

In (4b), we have $\Gamma^\alpha_{\nu\alpha} = (-g)^{-\frac{1}{2}} \partial\left[(-g)^{\frac{1}{2}}\right]/\partial x^\nu = g_{\alpha\beta,\nu} g^{\alpha\beta}/2$, so $\partial \Gamma^\alpha_{\nu\alpha}/\partial l = 2\partial(\chi_{,\nu}/\chi)/\partial l$ for the one term. (Here $g$ is the determinant of the metric tensor and a comma denotes the partial derivative.) The other term is $A^\alpha_{\nu;\alpha} = (-g)^{-\frac{1}{2}} \partial\left[(-g)^{\frac{1}{2}} A^\mu_\nu\right]/\partial x^\mu - (\partial g_{\alpha\beta}/\partial x^\nu) A^{\alpha\beta}/2$. We rewrite this using $(-g)^{-\frac{1}{2}} = \chi^2(-g^*)^{\frac{1}{2}}$ and $(-g)^{\frac{1}{2}} A^\mu_\nu = \chi\chi'(-g^*)^{\frac{1}{2}} \delta^\mu_\nu/2$. The resulting form may be simplified using $(-g^*)^{-\frac{1}{2}} \partial\left[(-g^*)^{\frac{1}{2}}\right]/\partial x^\nu = g^*_{\alpha\beta,\nu} g^{*\alpha\beta}/2$. The result is $A^\alpha_{\nu;\alpha} = (1/2)\partial(\chi_{,\nu}/\chi)/\partial l$. Setting the two terms equal gives $(1/2)(\chi_{,\nu}/\chi)' = 2(\chi_{,\nu}/\chi)'$ or $(\chi_{,\nu}/\chi)' = 0$. The solution is

$$l_0 = \text{constant} \quad , \tag{8}$$

in terms of the length in (7) above.

In (4c), we can evaluate $S_{\mu\nu}$ using (5) and previous relations. The vanishing of the spacetime components of the 5D Ricci tensor is then equivalent to saying that the 4D Ricci tensor is

$$^{(4)}R_{\mu\nu} = S_{\mu\nu} = \frac{3}{L^2}\left[\frac{l}{l-l_0}\right]^2 g_{\mu\nu} \quad . \tag{9}$$

Since an Einstein space as usually defined has $^{(4)}R_{\mu\nu} = \Lambda g_{\mu\nu}$, this relation also gives the effective cosmological constant for spacetime.

Let us now draw together our results. They involve a fiducial value $l_0$ of $x^4 = l$, a constant length $L$ and a function of the spacetime coordinates $k(x^\gamma)$. The last is arbitrary, so we can define a new metric tensor $\bar{g}_{\mu\nu} \equiv k(x^\gamma) g^*_{\mu\nu}(x^\gamma)$ in place of the one used above. Now from (9) we find that $^{(4)}R_{\mu\nu} = (3/L^2) k(x^\gamma) g^*_{\mu\nu}(x^\gamma) = (3/L^2) \bar{g}_{\mu\nu}$. Under a constant conformal transformation of the spacetime metric, $^{(4)}R_{\mu\nu}$ remains invariant; therefore, these are Einstein's equations (1) with cosmological constant $3/L^2$ for the metric tensor $\bar{g}_{\mu\nu}$. However, for every such 4D solution $\bar{g}_{\mu\nu}$ with cosmological constant $3/L^2$ of Einstein's equations, there is a corresponding 5D solution of the Ricci-flat field equations $R_{AB} = 0$ with line element

$$dS^2 = \left[\frac{l-l_0}{L}\right]^2 \bar{g}_{\mu\nu}(x^\gamma) dx^\mu dx^\nu - dl^2 \quad. \tag{10}$$

The 4D spacetime metric $g_{\mu\nu} = (1-l_0/l)^2 \bar{g}_{\mu\nu}(x^\gamma)$ has an effective cosmological "constant" given by

$$\Lambda = \frac{3}{L^2}\left[\frac{l}{l-l_0}\right]^2 \quad. \tag{11}$$

This 5D quantity goes back to the 4D one when $l_0 = 0$; but in general we have an embedding of 4D in 5D with a cosmological "constant" which depends on the value of the fifth coordinate.

3. Discussion

In the preceding section, we proved the

Theorem: Any solution of the 4D Einstein equations of general relativity in vacuum with a cosmological constant $3/L^2$ may be embedded in a solution of the 5D Ricci-flat equations with an effective cosmological "constant" given by $\Lambda = \left(3/L^2\right) l^2 / \left(l - l_0\right)^2$, where $l$ is the extra coordinate.

This has major implications for several physical situations, including: (a) our solar system as described by the Schwarzschild-de Sitter solution and other systems such as binary pulsars; (b) rapidly-spinning objects described by the Kerr metric, for which significant embeddings have been long sought; (c) gravitational waves, which are currently the subject of much research as for example by the Laser Interferometer Gravitational-Wave Observatory; (d) the early universe during its de Sitter-like phase of inflation.

To completely study these situations would take us far beyond the purview of the present work, because each involves a different solution of the Einstein equations. This is particularly true of the dynamics involved, which differs greatly from case to case. By contrast, the implications for the cosmological "constant" are generic. We therefore propose to give a summary of the dynamical effects of our embedding, drawing on results in the literature [11-14, 29-38], and follow it by a discussion of the possible behaviours for $\Lambda$. These two topics involve respectively the 5D line element (10) which contains a 4D Einstein one, and the corresponding cosmo-logical "constant" (11) which contains the Einstein one given by $3/L^2$. In those relations, $x^4 = l$ is the fifth coordinate as measured from some fiducial value $l_0$ and $L$ is a constant length. When $l_0 = 0$, (10) takes on the pure-canonical form, for which it is known that the 4D equations of motion take their standard form: the 4D motion is geodesic in the conventional sense, the Weak Equivalence Principle is obeyed and the 4D dynamics do not "know" about the embedding in 5D (see below). Similarly, when $l_0 = 0$, (11) gives $\Lambda = 3/L^2$ which is a true constant: this is commonly taken to involve a cosmological length of order $10^{28}$ cm, so $\Lambda$ is a small parameter which is identical to the standard Einstein one.

The situation changes significantly when $l_0 \neq 0$. While the form of the 5D metric is preserved for $l \to (l - l_0)$, the consequences for classical relativity are analogous to those of symmetry breaking in quantum field theory.

Dynamics is altered, notably by the appearance of a fifth force which acts in spacetime. This is a general consequence of any 5D metric whose 4D part depends on the extra coordinate (the pure-canonical form with a factor $l^2/L^2$ in front of a 4D part that is independent of $l$ is an exception). The existence of this force has been demonstrated in both induced-matter theory and membrane theory, the two currently popular versions of 5D relativity [31, 32]. It is not difficult to see why such a force arises, assuming that the dynamics is derived as usual by finding the extremum of the path length, in the 5D case via $\delta \left[ \int dS \right] = 0$. The result is a set of 5 equations, of which the first 4 describe motion in spacetime and the fifth describes motion in the extra dimension. However, in general these equations are <u>coupled</u>. For spacetime, this coupling may be expressed as a fifth force (per unit mass) or acceleration, which modifies conventional 4D geodesic motion. The fifth force involves the *l*-dependence of the 4D metric tensor, and the 'velocity' $dl/ds$ in the fifth dimension measured with respect to 4D proper time. If either of these vanishes, conventional geodesic motion is recovered. It is the independence of the 4D metric from $x^4 = l$ which is the basis for saying that the Weak Equivalence Principle can be viewed as a symmetry of the 5D metric [30]. This is bolstered by the fact that the second term in the fifth force $(dl/ds)$ is not in general zero [39]. The fact that this extra force depends on the velocity in the extra dimension shows that its origin is inertial in the Einstein sense (it depends on the velocity of 4D spacetime with respect to the 5D frame). For metrics which engender a fifth force, however small, we expect violations of the Weak Equivalence Principle. For metrics of the type we are discussing (with $l_0 \neq 0$), this extra force has been shown [14] to be

$$f^\mu = -\left[\frac{l_0}{l(l-l_0)}\right]\frac{dl}{ds}\frac{dx^\mu}{ds} \quad . \tag{12}$$

The explicit form of this in a given situation depends on $l = l(s)$ and $x^\mu = x^\mu(s)$. The first of these functions has been discussed elsewhere [14]. The second has also been discussed in special cases [6]; but the determination of $x^\mu(s)$ can be much simplified in the present case, because we can prove that the equation of motion in spacetime is equivalent to the geodesic equation for the metric tensor $\bar{g}_{\mu\nu}(x^\gamma)$ above.

To see this, let $d\sigma^2 = \bar{g}_{\mu\nu}(x^\gamma)dx^\mu dx^\nu$, which by (10) is related to the line element of spacetime via

$$\left(\frac{ds}{d\sigma}\right)^2 = \left(1 - \frac{l_0}{l}\right)^2 . \tag{13}$$

Using this with (12), there comes

$$\frac{d^2 x^\mu}{d\sigma^2} = \left(\frac{ds}{d\sigma}\right)^2 \left(\frac{d^2 x^\mu}{ds^2} - f^\mu\right) . \tag{14}$$

Moreover, the Christoffel symbols $\overline{\Gamma}^\mu_{\alpha\beta}$ associated with $\overline{g}_{\mu\nu}$ are such that $\overline{\Gamma}^\mu_{\alpha\beta} = \Gamma^\mu_{\alpha\beta}$, as the two metrics are related by a constant conformal transformation in spacetime; hence,

$$\overline{\Gamma}^\mu_{\alpha\beta} \frac{dx^\alpha}{d\sigma} \frac{dx^\beta}{d\sigma} = \left(\frac{ds}{d\sigma}\right)^2 \Gamma^\mu_{\alpha\beta} \frac{dx^\alpha}{ds} \frac{dx^\beta}{ds} . \tag{15}$$

Adding this to (14) shows that the 4D part of the equation of motion is equivalent to a geodesic worldline in the spacetime with $\overline{g}_{\mu\nu}$. Moreover, given $l = l(s)$ [14], we can use equation (13) to obtain $\sigma = \sigma(s)$. Finally, we can combine these results to obtain $x^\mu(s)$, and so evaluate the extra force (12) noted above.

That force obviously vanishes for $l_0 = 0$, in which case the 5D metric (10) becomes the pure-canonical one we have noted above as exceptional, and the cosmological constant (11) becomes the $3/L^2$ of standard general relativity. However, (12) does not depend explicitly on the constant $L$, confirming that it is an inertial force and not related to the cosmological-constant force (per unit mass) $\Lambda r/3$ typical of Einstein's theory. Since dynamics is known to be in good agreement with geodesic motion as based on general relativity, we infer that the extra force (12) is small in most situations. Nevertheless, it could in principle be detected, for example by the anomalous motions of galaxies [34], or by violations of the Weak Equivalence Principle in a proposed new test of this using artificial satellites [35].

The cosmological "constant" associated with a 5D embedding has properties which can differ greatly from the 4D situation. The latter is recovered for $l \gg l_0$ in (11), in which case the 5D metric (10) takes the pure-canonical form. Then, we could not tell from studies of the cosmological constant or dynamics if we were living in a 4D world or a 5D world (see above). However, there is no *a priori* reason for assuming anything about the relative sizes of the constants $L$, $l_0$ and the coordinate $x^4 = l = l(s)$. One consequence of this is that $\Lambda$ can diverge for $l \to l_0$, resulting in a self-consistent model of the inflationary universe in which $\Lambda$ is unbounded at the big bang and decays to its presently-observed value over cosmological time [14]. There are also possible applications of (10) and (11) to the cosmological-"constant" problem, insofar as there is no <u>unique</u> value of $\Lambda$. Rather, we have $\Lambda = \Lambda(l)$ with $l = l(s)$ in general. For example, if $l \ll l_0$ then the 5D metric (10) contains a 4D part which is a factor multiplied onto ordinary spacetime, and the cosmological constant (11) becomes $(3/L^2)(l/l_0)^2$. That is, the "standard" value is modulated by a factor that depends on the fifth dimension. (In this regard, it should be mentioned in passing that the "standard" value of the cosmological constant in 5D theory is reported in the literature by various workers as $3/L^2$ or $3/l^2$ for the pure-canonical metric, depending on whether the prefactor $(l/L)^2$ on the 4D metric is included or not: see note 40.) This variability of the cosmological "constant" can be investigated further by considering metrics of wider scope than what we have dealt with above.

Thus we can explicitly include a scalar field $\Phi = \Phi(x^\gamma, l)$ and both spacelike and timelike natures for the extra coordinate $(\varepsilon = \pm 1)$ by writing $dS^2 = g_{\alpha\beta}(x^\gamma, l) dx^\alpha dx^\beta + \varepsilon \Phi^2(x^\gamma, l) dl^2$. The motive for this is that the cosmological "constant" is widely regarded as measuring the value of a scalar field, the Einstein case corresponding to a perfect scalar fluid with effective density $\Lambda/8\pi$ and pressure $-\Lambda/8\pi$ (where the gravitational constant and the speed of light are unity). Then the field equations of general relativity read $G_{\mu\nu} = 8\pi T_{\mu\nu}$ in terms of the Einstein tensor and the energy-momentum tensor (where the latter includes the $\Lambda$-fluid). Now any solution of the 5D Ricci-flat equation $R_{AB} = 0$ may be regarded as a solution of the 4D Einstein equations if $T_{\mu\nu}$ has a certain form [7]. It is given by

$$8\pi T_{\alpha\beta} = \frac{\Phi_{,\alpha;\beta}}{\Phi} - \frac{\varepsilon}{2\Phi^2} \left\{ \frac{\Phi_{,4} g_{\alpha\beta,4}}{\Phi} - g_{\alpha\beta,44} + g^{\lambda\mu} g_{\alpha\lambda,4} g_{\beta\mu,4} \right.$$

$$\left. - \frac{g^{\mu\nu} g_{\mu\nu,4} g_{\alpha\beta,4}}{2} + \frac{g_{\alpha\beta}}{4} \left[ g^{\mu\nu}_{,4} g_{\mu\nu,4} + \left(g^{\mu\nu} g_{\mu\nu,4}\right)^2 \right] \right\} \quad . \tag{16}$$

This is a combination of "ordinary" matter and scalar-field matter, where the energy density of the latter implicitly determines a "local" value of the cosmological "constant" that depends on the fifth dimension.

Even for the modest embedding which we have discussed it is clear that the fifth dimension can affect local 4D physics. By (10) and (11), the behaviour of the fifth coordinate can ex-pand or shrink spacetime and change the value of the effective cosmological "constant". Detailed analysis of these things should be carried out to see if the world has more than 4 dimensions.

Acknowledgements

Thanks for past comments go to H. Liu and S.S. Seahra. This work was supported in part by N.S.E.R.C. of Canada and the University of Missouri at Columbia.

39. On $dl/ds \neq 0$: This may be proved by taking the $l$-component of the 5D geodesic, along with the usual normalization condition for the 4-velocities $u^\alpha u_\alpha = 1$, and proving that $dl/ds = 0$ leads to a contradiction. (See ref. 6, p. 72.)

40. On $\Lambda = 3/L^2$ versus $3/l^2$: For the canonical metric, there is a ready reduction of the 5D field equations to the 4D ones with an effective cosmological constant which is related to the 4D Ricci scalar via $^{(4)}R = 4\Lambda$. For a spacelike extra dimension $\Lambda > 0$ while for a timelike one $\Lambda < 0$. However, the size of $\Lambda$ depends on whether one considers the 4D part of the 5D space with or without the prefactor $(l/L)^2$. This affects $^{(4)}R$ and so $\Lambda$, and explains why in the literature there is reported both $\Lambda = 3/L^2$ and $\Lambda = 3/l^2$. Both choices are valid, from an algebraic viewpoint. To see this, consider the metric for 3D Euclidean space in spherical polar coordinates: $dr^2 + r^2(d\theta^2 + \sin^2\theta d\phi^2)$. For a 2D spherical section, $r$ is often suppressed by putting it equal to unity, but clearly this is not justified if one can go outside the 2-surface. From a physical viewpoint, the choice between the two forms for $\Lambda$ depends on what measurements determine how 4D spacetime is embedded in a 5D manifold. (See ref. 6, p. 140.)